\documentclass[conference]{IEEEtran}
\usepackage{cite} 
\usepackage{amsmath,amssymb,amsfonts}
\usepackage{graphicx}
\usepackage{textcomp}
\usepackage{xcolor}
\def\BibTeX{{\rm B\kern-.05em{\sc i\kern-.025em b}\kern-.08em
    T\kern-.1667em\lower.7ex\hbox{E}\kern-.125emX}}

\usepackage{listings}
\usepackage{hyperref}
\usepackage{booktabs} 
\usepackage{wrapfig}
\usepackage{fancyvrb}
\usepackage{setspace}
\usepackage{mdwlist}
\usepackage{boxedminipage}
\usepackage{textcomp}
\usepackage{color}
\usepackage{verbatim}
\usepackage{flafter}
\usepackage{listings}
\usepackage{amsmath}
\usepackage{graphicx}
\usepackage{caption}
\usepackage{subcaption}  
\usepackage[figuresright]{rotating}
\usepackage{epstopdf} 
\usepackage{placeins}
\usepackage[T1]{fontenc}  
\usepackage{color, colortbl}
\usepackage{multirow}
\usepackage{algorithm}
\usepackage{algpseudocode}
\usepackage{amssymb}

\begin{document}

\title{A Real-Time, Auto-Regression Method for In-Situ Feature Extraction in Hydrodynamics Simulations}

\author{\IEEEauthorblockN{1\textsuperscript{st} Kewei Yan}
\IEEEauthorblockA{\textit{Department of Computer Science} \\
\textit{University of North Carolina at Charlotte}\\
Charlotte, USA \\
kyan2@charlotte.edu}
\and
\IEEEauthorblockN{2\textsuperscript{nd} Yonghong Yan}
\IEEEauthorblockA{\textit{Department of Computer Science} \\
\textit{University of North Carolina at Charlotte}\\
Charlotte, USA \\
yyan7@charlotte.edu}
}

\maketitle

\begin{abstract}
Hydrodynamics simulations are powerful tools for studying fluid behavior under physical forces, enabling extraction of features that reveal key flow characteristics. Traditional post-analysis methods offer high accuracy but incur significant computational and I/O costs. In contrast, in-situ methods reduce data movement by analyzing data during the simulation, yet often compromise either accuracy or performance.
We propose a lightweight auto-regression algorithm for real-time in-situ feature extraction. It applies curve-fitting to temporal and spatial data, reducing data volume and minimizing simulation overhead. The model is trained incrementally using mini-batches, ensuring responsiveness and low computational cost.
To facilitate adoption, we provide a flexible library with simple APIs for easy integration into existing workflows. We evaluate the method on simulations of material deformation and white dwarf (WD) mergers, extracting features such as shock propagation and delay-time distribution. Results show high accuracy (94.44\%–99.60\%) and low performance impact (0.11\%–4.95\%), demonstrating the method’s effectiveness for accurate and efficient in-situ analysis.

\end{abstract}


\section{Introduction}
\label{sec:introduction}
Hydrodynamics simulations have become a powerful computational tool for analyzing fluid behavior.
These simulations enable scientists to explore various conditions and scenarios, deepening their understanding of fluid interactions and dynamics~\cite{sorokin2004analysis}.
To enhance their comprehension of fluid behavior, researchers utilize feature extraction techniques to correlate these behaviors with key variables in hydrodynamics simulations~\cite{javanmard2019hydrodynamic}.
Based on established correlations, feature extraction can directly identify hydrodynamic phenomena in relation to specific variables.
For example, Chen et al.~\cite{chen2012predicting} predicted storm surges by analyzing water level distribution features under varying surface conditions.
In other scenarios, feature extraction identifies specific variables linked to targeted hydrodynamic phenomena.
For example, Chinta et al.~\cite{chinta2024machine} pinpointed key parameters influencing wetland methane emissions by isolating features associated with methane production, oxidation, and transport.
These advancements not only deepen our knowledge of fluid dynamics but also inform practical solutions to real-world challenges~\cite{hossain2022high}.

Despite significant advancements in the field, the methodologies employed in feature extraction remain a key consideration.
As a data-driven approach, feature extraction methods are aiming at achieving high accuracy while minimizing the overhead to the simulation~\cite{moon2020classification}.
However, a persistent trade-off has been observed: high-accuracy methods remain computationally expensive, while low-overhead alternatives lack the precision required for comprehensive fluid analysis~\cite{varga2013towards}.
%
Traditional methods often prioritize accuracy and rely heavily on post-analysis processing to maximize the use of simulated data.
As modern hydrodynamics simulations grow in complexity and scale, these input/output (I/O) requirements can become significant bottlenecks~\cite{khan2022hvac}. 
Additionally, tasks like organizing and transforming large datasets can lead to increased complexity and inconvenience~\cite{li2020elasticbroker}.
In contrast, in-situ methods conduct analysis directly within the simulation environment significantly reducing I/O overhead, effectively addressing one of the major limitations of traditional post-analysis methods~\cite{qamar2022deep}.
However, these methods continue to face challenges in balancing accuracy with their impact on simulation performance.
Overly complex algorithms can introduce significant computational overhead~\cite{mangalam2004real}, while simplified approaches risk compromising the quality of feature extraction and analysis~\cite{padmanabha2018mitigating}.

In this paper, we present a novel real-time, auto-regression-based in-situ feature extraction method for hydrodynamics simulations.
Our method effectively reduces computational demands while delivering high-accuracy feature analysis, thereby efficiently providing deeper insights into fluid behavior.
The highlights and contributions of our method are as follows:
1) We present an in-situ feature extraction approach that collects data points across both temporal and spatial dimensions during simulations, leveraging auto-regressive models to fit and predict the distribution of targeted features.
This dual-dimensional strategy enables users to extract features from partial data, eliminating the need for large-volume datasets and facilitating both rapid and comprehensive data analysis.
2) We employ a linear auto-regressive model for in-situ analysis, and train the model with mini-batches of generated data during simulation.
This method requires low computational demands for modeling, and enables real-time curve fitting and predictions of targeted feature distributions for feature extraction.
3) We implement a library framework that offers users simple APIs for programming feature extraction methods tailored to their specific features and models.
This flexibility enhances adaptability and boosts user productivity by adopting our methods to hydrodynamics simulations.

The effectiveness of this method is demonstrated through its application to material deformation analysis and white dwarf (WD) detonation determination.
We apply our approach to LULESH~\cite{LULESH2:changes} for the analysis of material deformation.
We extract velocity distribution cross domain from LULESH, and gain valuable insights into the pattern of blast wave propagation through materials, for example, a rapid drop of velocity value during early stages.
For WD detonation determination, we apply our approach to the Castro wdmerger simulation~\cite{almgren2020castro}.
Four key features were extracted: temperature, angular momentum, mass, and energy distributions.
These enable us to observe relationships between the delay-time distribution (DTD) of supernovae and various astrophysical events, such as mass ejection.
The results also demonstrate the method’s effectiveness in enhancing in-situ analysis, with high feature analysis accuracy ranging from 94.44\% to 99.60\%, and minimal performance impact of just 0.11\% to 4.95\%.

In the rest of this paper, we present background and motivation in section~\ref{sec:background}.
We discuss detailed methodology in section~\ref{sec:methodology}, and we show evaluations on LULESH and Castro wdmerger in sections~\ref{sec:lulesh} and~\ref{sec:castro}. 
We finally include related work in section~\ref{sec:related_work} and conclude this paper in section~\ref{sec:conclusion}.

\section{Background and Motivation}
\label{sec:background}
Hydrodynamic features are usually a function of key variables involved in hydrodynamics simulation, which can include observable patterns, structures, or metrics that indicate how fluids behave.
Scientists employ feature extraction to correlate fluid behaviors with key
variables in hydrodynamics simulations~\cite{javanmard2019hydrodynamic}.
These techniques enable scientists to gain deeper insights into hydrodynamic phenomena~\cite{mangalam2004real}, enhancing their ability to predict fluid behavior~\cite{kempf1990visualizing}, understand instabilities~\cite{gupta2022advances}, and model complex systems~\cite{wolf2020shape}.
For example, as a function of fluid density, velocity, and viscosity, the Reynolds number is commonly extracted as a feature serving as a critical indicator of flow characteristics~\cite{rott1990note}.

Feature extraction involves collecting data by selecting subsets of simulation results within specific temporal or spatial domains to focus on regions of interest (ROI) and modeling this data to convert key hydrodynamic variables into meaningful features.
For example, in non-destructive testing (NDT)~\cite{blitz1997electrical}, electromotive force is measured as a key variable to accurately identify areas of weakness within materials.
During the simulation, the electromotive force data crossing the whole domain are continuously stored.
Based on the relationship between resistance and the complete electromotive force data stored from simulation, this feature extraction enables the identification and visualization of regions with low resistance, effectively highlighting areas of compromised strength.
Despite a direct identification described above, the obtained features also contributes to in-depth analysis, such as threshold detection~\cite{hall2017identifying} and regression analysis~\cite{mesyats2013hydrodynamics}.

For another example, Cheng et al.~\cite{cheng1991interfacing} accurately characterized tidal hydrodynamic phenomena through post-analysis by integrating all relevant components generated by the simulation.
While traditional methods achieve a solid understanding of hydrodynamic phenomena with high accuracy, they often place heavy demands on data processing.
As noted by Sodani et al.~\cite{sodani2011race}, large-scale simulations can generate large amonut of data in a short period of time (between 200 and 300 PB/s in memory), posing a significant challenge for storage and I/O operations in data transport~\cite{hamilton2017extreme}.
Additionally, post-analysis involves essential but complex tasks such as organizing and transforming large datasets, which further complicate the process triggering inconvenience~\cite{li2020elasticbroker}:
As distributed systems are widely utilized to handle data-intensive, large-scale hydrodynamics simulations~\cite{hui1999hydrodynamic}, additional factors such as data partitioning, allocation, access, synchronization, and system stability must be carefully considered~\cite{destanoglu2008randomized}.

In contrast, in-situ methods, which perform feature extraction during the simulation process, effectively address several limitations of post-analysis approaches, such as the challenges of extensive data transport~\cite{ma2009situ}.
By eliminating the need for large-scale data transfers, these methods significantly lower I/O operations, which are often a critical performance bottleneck in post-analysis processing~\cite{khan2022hvac}.
However, in-situ analysis methods face persistent challenges on maintaining accuracy without significantly impacting simulation performance~\cite{ayachit2016performance}.
For example, in-situ visualization of large datasets can be highly time-consuming due to the resource-intensive rendering process, potentially leading to interruptions in the simulation~\cite{kress2019comparing}.
In-situ methods that rely on reduced datasets~\cite{wright2017improving} or simplified models~\cite{nicholas2012modelling} can achieve faster execution speeds with minimal impact on simulation performance, but often compromise the accuracy of feature analysis~\cite{duan2020data}.

As scientists delve deeper into complex fluid behaviors, it becomes clear that the efficiency of these methods is just as important as their accuracy~\cite{chen2012enabling}.
Several solutions are presented to balance high accuracy with low computational overhead in feature extraction.
To enhance accuracy, some methods employ advanced functions to construct hydrodynamic features from key variables in simulations, optimizing the use of datasets.
For example, Seitenzahl et al.\cite{seitenzahl2009spontaneous} utilized various temperature change formations to predict the detonation of white dwarfs.
However, we observe that features extracted using these methods can still result in low accuracy in downstream analyses\cite{wang2019detecting}, primarily due to the limited information utilized and the inherent uncertainty in the correlation between the features and hydrodynamic phenomena~\cite{chinta2024machine}.
Our approach leverages both temporal and spatial dimensions to enhance the richness of information extracted from datasets.
Additionally, by employing an auto-regressive model for curve-fitting, our method eliminates the need to process large volumes of evolving data thus reduce the execution time of data processing.

To minimize overhead, learning-based models have been integrated into in-situ analysis to facilitate real-time feature extraction from simulated data~\cite{yang2023lightweight}.
However, these methods often require extensive training time and large amounts of training data to effectively pre-train the models~\cite{wang2017understanding}.
Also, these methods are historical data-driven, thus have limited capability in handling time-sensitive tasks such as weather forecasting~\cite{hu2020real} and flood prediction~\cite{ming2020real}, thereby constraining their utility.
Our approach avoid pre-training, and train the model with mini-batch as the data is simulated during simulation to enable real-time feature extraction without any constrains on training.

\section{Methodology and Implementation}
\label{sec:methodology}

\subsection{Auto-Regression Algorithm}
An auto-regressive model is a statistical approach that describes the relationship of a variable with its own past values across different temporal or spatial characteristics.

A typical linear format can be shown as:
\[V(l,t) = \beta_0 + \beta_1 V(l-1,t-lag) + ... + \beta_n V(l-n,t-lag) + \epsilon\]

Where, $V$ is the targeted variable of simulation, $n$ is model size, $l$ is the location, and $t$ is the time step.
$V(l,t)$ represents the variable with temporal and spatial characteristics.
$lag$ is the time step lag, which is measured by number of iterations.
$\beta$'s are the regression coefficients, and $\epsilon$ is the error term.

Like ordinary linear models, optimization methods such as gradient descent (GD)~\cite{wilms2018combining} are utilized during training to minimize prediction error and estimate the parameters.
Once the model is effectively trained, it can generate predicted values for any specified set of past values with temporal and spatial characteristics, leveraging the learned relationships.

The predictive relationship is represented as:
\[V(l,t) = b_0 + b_1 V(l-1,t) + ... + b_n V(l-n,t)\]

Where, {$b_0$, $b_1$, $b_2$, $b_3$, ..., $b_n$} are the trained coefficients, and a prediction of $V(l,t)$ involves $n$ past values. To forward the targeted variable cross time and space, we replace the $V(l,t)$ by $V(l+1,t)$ and $V(l,t+1)$, respectively.

\subsection{Real-Time and Auto-Regression Based In-Situ Feature Extraction}
Real-time in-situ feature extraction consists of three primary components: data collection, curve-fitting, and variable tracking. 
Leveraging the iterative structure of hydrodynamics simulations~\cite{livne1993implicit}, for data collection, our approach continuously captures data points based on predefined temporal and spatial characteristics as they are generated during simulation iterations.
For curve-fitting using these collected data points, a linear auto-regressive model is trained on mini-batches in parallel with the simulation, allowing efficient curve-fitting for extracting meaningful features.

\subsubsection{Data Collection}
Real-time data collection has been widely integrated into scientific research.
For example, Dal et al.~\cite{dal2015single} introduced a Single Column Model (SCM) for climate simulations, which collects diagnostic variables, such as the total mass of the water column, every few hours to provide real-time insights into climate change trends.
However, focusing on a single point can obscure important spatial patterns, such as gradients~\cite{fasullo2023overview}.
To address this, we incorporate spatial characteristics into one-dimensional real-time data collection, enhancing flexibility and increasing the informativeness of the collected data.
In our approach, data points are gathered at each iteration of the simulation.
A helper function continuously monitors each iteration for the specified temporal and spatial characteristics, ensuring that data collection aligns with the user-defined parameters.
When the defined conditions are met, the helper function efficiently aggregates the relevant data into mini-batches.
By doing so, we facilitate more accurate feature extraction and analysis, enabling deeper insights into the hydrodynamic phenomena being studied.

\subsubsection{Curve-Fitting}
The auto-regression algorithm is notable for its ability to correlate the current and future values of a variable with its past values. This feature makes it particularly effective for curve-fitting applications, such as modeling the evolution of force~\cite{pang2021method} and energy~\cite{ni2021numerical} during wave propagation.
By leveraging this capability, we employ an auto-regressive model to capture the evolution of key variables for comprehensive feature extraction.
This approach allows us to effectively analyze dynamic systems in real time.
To maximize the benefits of real-time data collection, we integrate the training loop directly into the simulation process for continuous curve-fitting.
As each mini-batch is filled with data, the model's parameters are immediately updated using GD within the current iteration.
After the update, the mini-batch is reset to collect new data for the next round of parameter adjustments, while the GD algorithm remains idle, awaiting the next filled mini-batch.
Once data collection concludes, the trained model is employed to perform inference, predicting the values of diagnostic variables at specified target locations.
During this process, the predicted values are ready for feature extraction, for example, to obtain the distribution of outliers, these values can be compared against predefined thresholds to assess their significance, if a predicted value does not exceed the threshold, the location is adjusted by a specified radius, enabling a more refined search for critical data points.

\subsubsection{Variable Tracking}
\begin{figure}[htb!]
 \centering
  \includegraphics[width=0.8\linewidth]{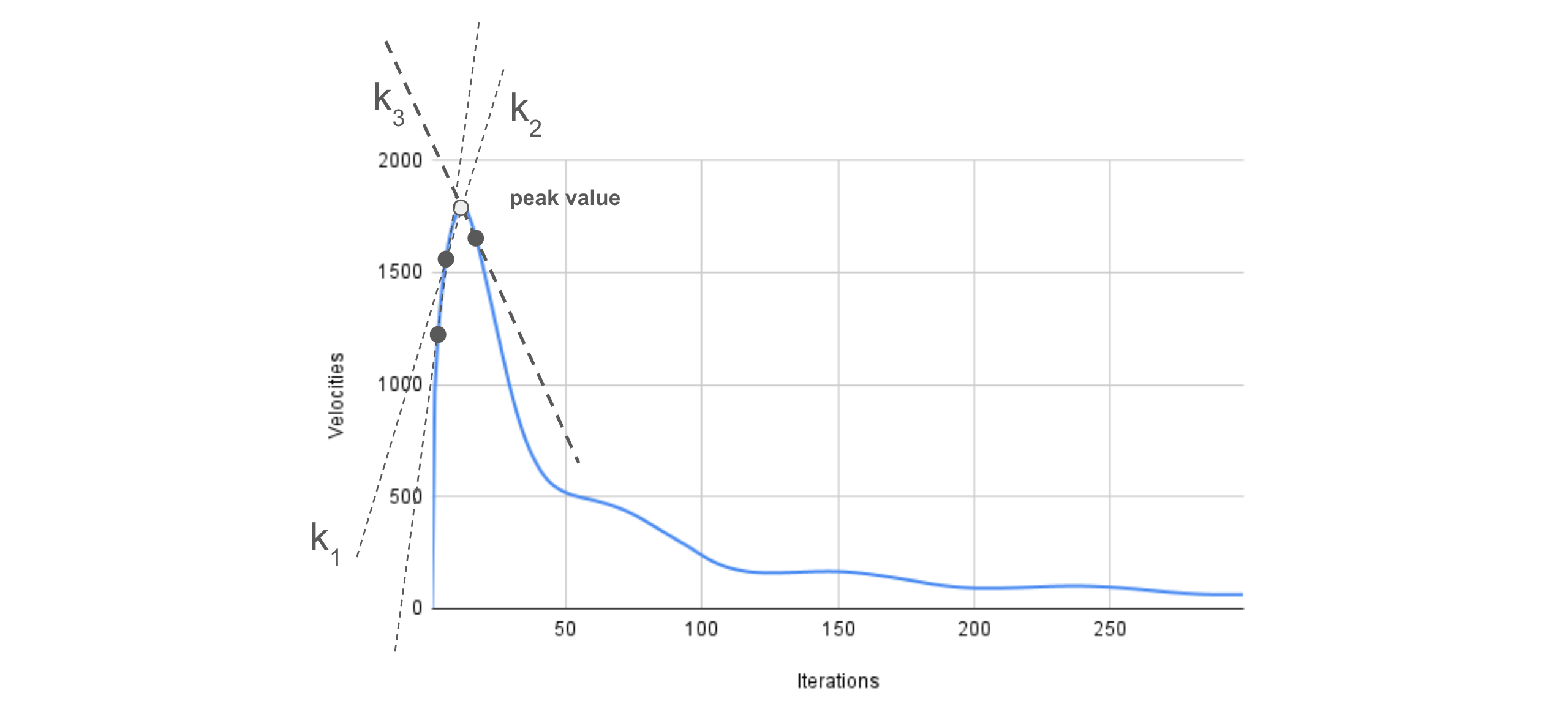}
 \caption{Illustration of variable tracking and peak value detection. The gradients such as $k_1$, $k_2$, and $k_3$ are computed, and in this case the negative $k_3$ indicates the peak value is detected.}
 \label{fig:var_track}
\end{figure}

Variable tracking is an algorithm to locate the focal points on a curve, such as local maxima/minima, and inflection points. 
The illustration of the basic algorithm locating local maxima is shown in Figure \ref{fig:var_track}.
Since each iteration represents a simulation time step, the acceleration, i.e. gradient, can be computed by subtracting the previous iteration's velocity from the current velocity.

We compute the gradients $k_1$, $k_2$, and $k_3$ by four back-to-back velocities. Once a positive $k_2$ while negative $k_3$ is observed, the velocity sampled from former iteration generating $k_3$ is evaluated as local maximum. 
Instead, suppose that the negative $k_2$ while positive $k_3$ is observed, the corresponding velocity is evaluated as local minimal.
Additionally, detecting local maxima in the derivative of the data enables precise identification of inflection points.
Through curve-fitting, our approach applies variable tracking to extract distribution features, identified by local maxima or inflection points, enabling efficient handling of threshold-based feature extraction tasks.

\subsection{API Design of the Library Framework}
To support the adoption of our approach, we provide a flexible library framework with simple APIs to help adopting our methods in a more productive way.
We designed two groups of APIs for the definition and execution of the feature extraction.
For definition, we designed three APIs.
\texttt{td\_var\_provider} is provided for users implementing the function accessing to the value of diagnostic variables.
The prototype has been pre-defined and the inputs are domain object of the simulation and location.
\texttt{td\_item\_para\_init} is used for initializing temporal and spatial characteristics of real-time data collection, and the input is formatted as tuple of three elements, for begin, end, and steps.
\texttt{td\_region\_add\_analysis} is used to construct the object of data analysis taking advantage of the presets of variable provider and temporal and spatial characteristics.
Besides, the name of feature extraction method, the feature to be extracted, and the flag indicating actions when the data analysis is exited are needed.
Till now, the framework supports threshold-based feature extraction, and methods of 'Curve\_Fitting' for data analysis.
For execution, we designed another three APIs.
\texttt{td\_region\_init} is used to initialize the object of analyzer.
\texttt{td\_region\_begin} and \texttt{td\_region\_end} are paired functions used to mark the start and end of code blocks for feature extraction.
In addition to executing the auto-regression algorithm, these callbacks also monitor the temporal and spatial characteristics, broadcasting values such as the current predicted value, the MPI rank indicating the location of the wave front, and a flag indicating the actions taken after the feature extraction process concludes.

\subsection{Code Example with LULESH}
We use LULESH code to demonstrate how to implement feature extraction using the APIs provided by our library framework.
The code, which includes the APIs and the main computations, is illustrated in Figure \ref{fig:lulesh_api}.
To program the feature extraction, users need to initial the feature extraction (line 10), implement a provider which accesses velocity by location id (lines 1 to 5), define the temporal and spatial characteristics of velocities by providing tuples of three (lines 11 and 12), and create an analyzer by specifying the method of data analysis as 'Curve\_Fitting' (lines 13 to 20).
In addition to the inputs mentioned above, extra parameters, such as a threshold (line 14) and a flag (line 15), are included in the analyzer.
These enable threshold-based feature extraction and manage post-analysis actions once data processing is complete.
At last, users just need to find out the code blocks of the main computation (line 25 and 26), and insert the lines indicating the begin (line 23) and end (line 28) of execution of data analysis, i.e. curve-fitting, to the original code.

\begin{figure}[h]
\centering
\begin{lstlisting}
double td_var_provider(Domain *locDom, int loc) {
  double v = locDom->xd(loc);
  
  return v;
}

int main (...) {
  ...
  // init td_region
  td_region_t *lulesh_region = td_region_init("", locDom);
  td_iter_param_t *lulesh_loc = td_iter_param_init(6, 10, 1);
  td_iter_param_t *lulesh_iter = td_iter_param_init(50, 373, 10);
  int method = Curve_Fitting;
  double threshold = 25.26;
  int if_simulation_will_terminate = 0;
  td_region_add_analysis(lulesh_region, \
                         td_var_provider, lulesh_loc, \
                         method, lulesh_iter, \
                         threshold, \
                         if_simulation_will_terminate);
  ...
  while (...){
    td_region_begin(lulesh_region);

    TimeIncrement(*locDom); //time-step update
    LagrangeLeapFrog(*locDom); //main computation

    td_region_end(lulesh_region);
  }
  ...
}
\end{lstlisting}
  \caption{Code to show the usage of APIs from our library framework to program feature extraction. The main computation is shown in lines 25 and 26. Bold functions are provided by the library framework.}
  \label{fig:lulesh_api}
\end{figure}


\subsection{Experiment Environment Settings}
We tested our approach on the server which is equipped with 2 Intel Xeon Gold 6230N CPUs for total 40 cores, 512G DRAM(4x64GB DDR4-2933), 4x128G Intel DC PMEM, 360GB Intel 3DXPoint DC P4800, Intel DC P4610 1.6TB NVMe.
In Section \ref{sec:lulesh} and Section \ref{sec:castro}, we provide a detailed evaluation of the accuracy and computational overhead of our real-time, auto-regression-based in-situ feature extraction method as applied to material deformation analysis with LULESH~\cite{LULESH2:changes} and white dwarf (WD) detonation determination with Castro’s wdmerger simulation~\cite{almgren2020castro}.

\section{Case 1: Material Deformations Analysis}
\label{sec:lulesh}
Material deformation analysis is essential for understanding how materials respond to external forces, providing insights into structural stability and safety~\cite{hu2004characterization}.
This is particularly crucial for extreme conditions, where identifying failure thresholds (break-points) informs design and safety measures~\cite{hasan2013effects}.
By analyzing force-induced wave propagation, we can define safe zones where particle velocities remain low enough to minimize damage~\cite{wang2017identification}.

\begin{figure}[ht]
  \centering
  \includegraphics[width=0.5\linewidth]{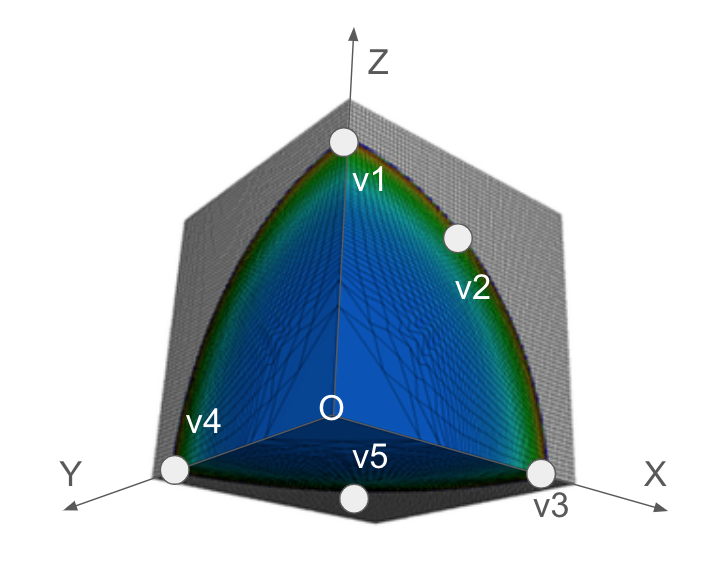}  
    \caption{Illustration of blast wave propagation within a 3D domain. The wave originates at point O, with velocities at any position (x, y, z) computed based on their components along each axis. The arc surface represents the wavefront, where velocities (v1, v2, ..., v5) are equal due to spherical symmetry. This illustration is adapted from~\cite{hornung2011hydrodynamics}.}
  \label{fig:LULESH_area}
\end{figure}

We use the LULESH mini-application to simulate material deformation under extreme conditions.
LULESH models wave propagation from a Sedov blast, assuming uniform initial energy within a cubic domain.
This setup captures essential wave dynamics like spherical symmetry while being computationally efficient.
An illustration of the blast wave propagation is shown in Figure \ref{fig:LULESH_area}, where waves spread evenly across x-, y-, and z-axes, and velocities on the same arc (e.g., $v1$, $v2$, ..., $v5$) share identical values.
LULESH employs Leapfrog timestepping to track velocities at each timestep, capturing material displacement over time and space. 
Mesh elements are updated iteratively, accounting for interactions with neighboring elements, enabling precise modeling of material behavior under intense forces.

In our study, we applied LULESH to domains of sizes 30, 60, and 90, examining velocity as the primary variable indicative of material motion.
To identify a ROI within which material motion falls below a critical safety threshold, we defined a series of velocity thresholds, ranging from 0.1\% to 20\% of the velocity initiated by the blast.
This approach enabled us to capture the radius of the ROI, revealing the extent of safe motion zones over time.
We predict these ROIs in real-time, implementing early termination of the simulation once the model reached a predefined accuracy threshold.
This early termination reduces computational workload substantially while maintaining accurate detection of the material break-points.

Additionally, we evaluated the performance impact of this variable tracking approach by running LULESH with 1, 8, and 27 MPI processes and measuring the overhead associated with MPI broadcasting required to keep all processes updated on the threshold detection status.
Curve-fitting was performed on datasets from various stages of the simulation (40\%, 60\%, and 80\% of total iterations), confirming that our method accurately captures the material’s response patterns and allows for efficient, early termination.
By balancing accuracy with performance, our approach addresses the computational challenges typical of large-scale hydrodynamics simulations, offering a reliable solution for feature extraction while minimizing overhead.

\subsection{Accuracy of Material Deformations Analysis}
We evaluated curve-fitting on two-dimensional data, incorporating both temporal and spatial characteristics, to calibrate the auto-regressive model for improved prediction accuracy.
As shown in Table \ref{tab:pred-1}, we observed instances of overfitting when data within the interval (10, 20), which are located centrally in the domain, was applied in the early stages.
Similarly, lower accuracies were noted when using data from the interval (20, 30) in the right-side region of the domain, also during early stages.
Based on these error distribution patterns, we selected data points on the left side, and collect them in early stages of simulation for further refinement of the curve-fitting process.

\begin{table}[!htb]
\resizebox{\linewidth}{!}{%
  \begin{tabular}{|p{0.3\linewidth}|p{0.2\linewidth}|p{0.2\linewidth}|p{0.2\linewidth}|}
  \hline
  \multirow{2}{0.3\linewidth}{\textbf{Locations}} & \multicolumn{3}{|c|}{\textbf{Ratio of Total Number of Iterations (\%)}} \\ \cline{2-4}
  & \textbf{40} & \textbf{60} & \textbf{80} \\ \hline
  (1, 10)   & 6.5\%  & 6.4\%  &  1.8\%  \\ \hline
  (10, 20)  & 267\%  & 3.9\%  &  3.1\%  \\ \hline
  (20, 30)  & 53.5\% & 80.4\% &  3.1\%  \\ \hline
  \end{tabular}%
 }
\caption{Error rates of curve-fitting (\%) for velocity as the diagnostic variable, using training data from 40\%, 60\%, and 80\% of total iterations with a domain size of 30.}
\label{tab:pred-1}
\end{table}

\begin{figure}[ht]
  \centering
  \includegraphics[width=0.8\linewidth]{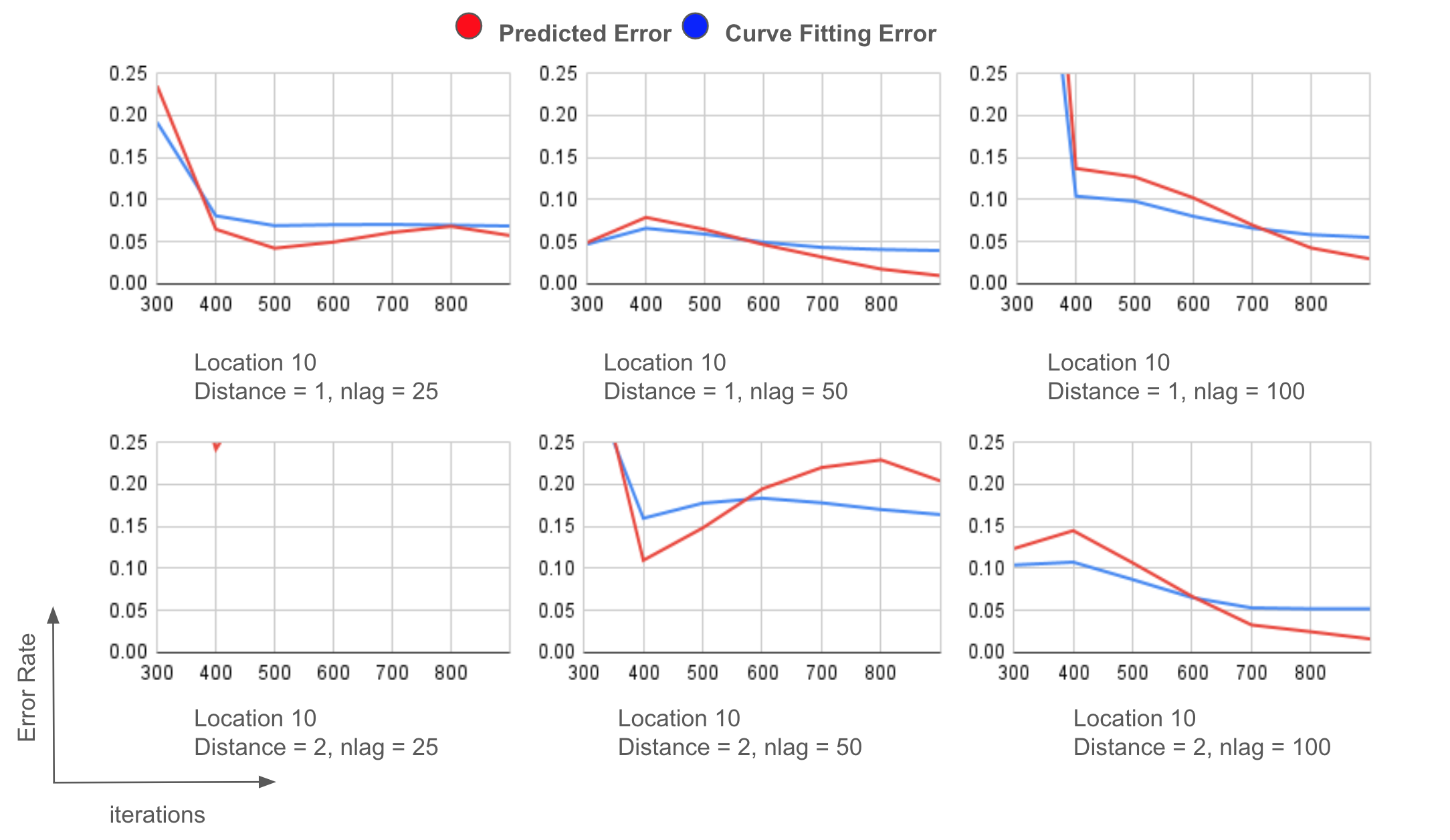}  
    \caption{Curve-fitting results on velocities at location 10 using lag values of 50 and 100 over 40\%, 60\%, and 80\% of total iterations with a domain size of 30. Only error rates between 0.00 and 0.25 are shown in the figures.}
  \label{fig:reg_fine_tune}
\end{figure}

To fine-tune the lag parameter in the auto-regressive model, we considered both temporal and spatial characteristics comprehensively. As illustrated in Figure \ref{fig:reg_fine_tune}, we applied a lag of 50 to optimize the model's performance for subsequent analyses. This choice was based on data from locations 1 to 10, using the first 40\% of the simulation iterations.

\begin{figure}[ht]
  \centering
  \includegraphics[width=0.8\linewidth]{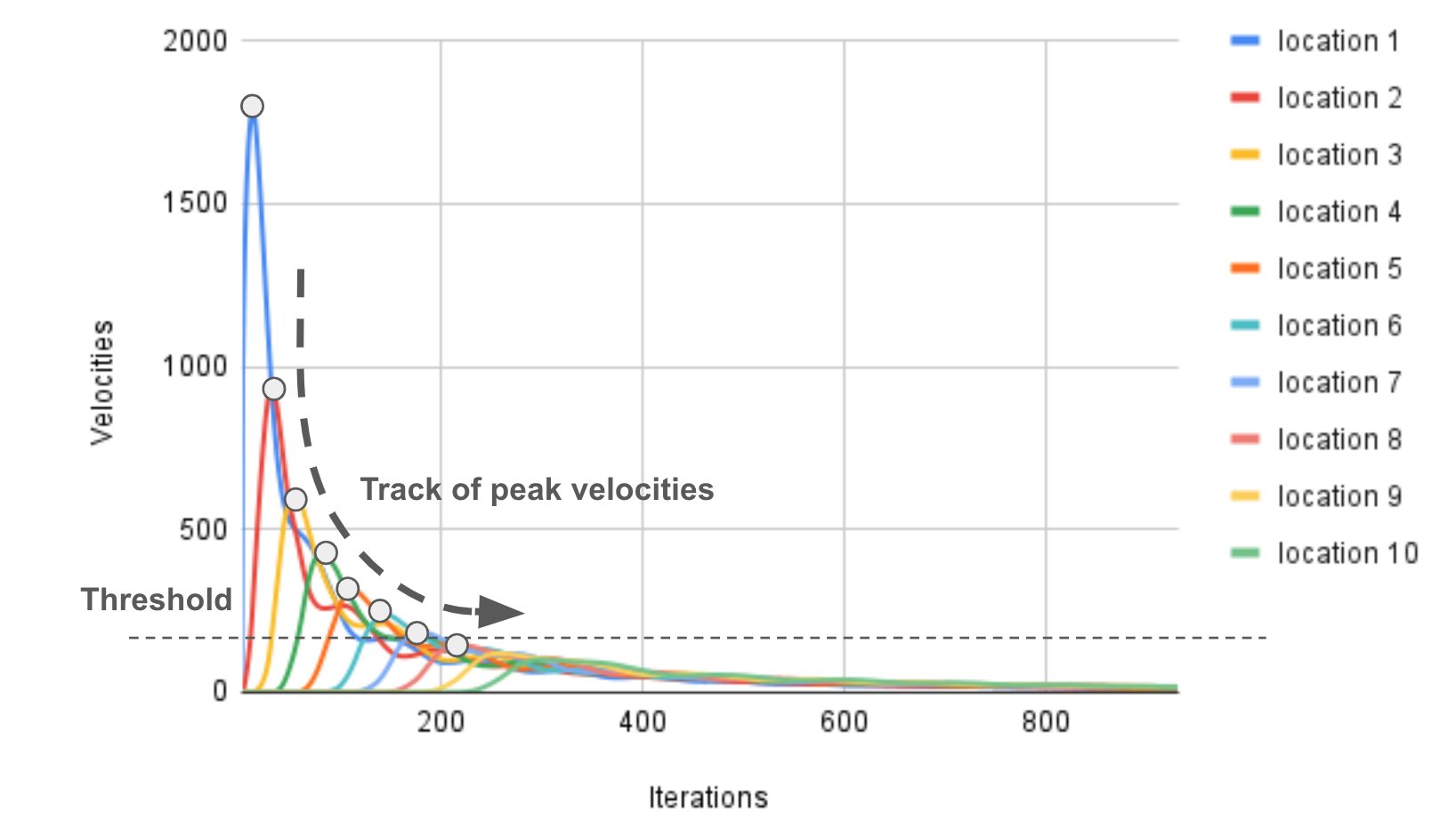}  
    \caption{Distribution of velocity over timesteps 1 to 932 at locations 1 to 10. The velocity curves are aligned into the same figure by number of iterations. The threshold is used to determine the breaking-point of certain material as the feature to be extracted.}
  \label{fig:LULESH_velocity}
\end{figure}

The Table \ref{tab:accuacy-1} shows the radius of the material break-point determined using feature extraction, compared to the 'ground truth' values from the simulation across a range of velocity thresholds in a domain of size 30.
Notably, at lower thresholds (0.1\% to 1\%), a consistent error of -16.67\% is observed, while at higher thresholds (5\% to 20\%), the two methods align closely, with minimal or zero difference. This indicates that the accuracy of feature extraction improves with increasing threshold levels.
This result demonstrates that our approach effectively extracts break-point features within the specified velocity thresholds, which range from 2\% to 20\% of the initial velocity. For a domain size of 30, our method successfully identifies break-points that cover a radius of 22 out of 30 units, representing approximately 53.7\% of the region.
As illustrated in Figure \ref{fig:LULESH_velocity}, we observe that the peak velocity of material motion decreases with increasing radius, indicating wave attenuation over distance.
Within this central 53.7\% area, the majority of the propagation patterns can be captured, such as the severely drop of velocity in early stages of the simulation.
For larger domain sizes, such as 60 and 90 cubic units, our method captures regions with radii of 51 and 76 units, covering approximately 72.3\% and 71.3\% of the total region, respectively.
This enhanced coverage contributes to deeper insights into material deformation analysis.

\begin{table}[!htb]
 \resizebox{\linewidth}{!}{%
  \begin{tabular}{|p{0.3\linewidth}|p{0.3\linewidth}|p{0.3\linewidth}|p{0.3\linewidth}|}
  \hline
    \textbf{Threshold(\%)} & \textbf{From Sim.} & \textbf{Feat. Extraction} & \textbf{Difference(\%)}\\ \hline
    0.1  & 25 & 30 & -5(-16.67\%)  \\ \hline
    0.2  & 25 & 30 & -5(-16.67\%)  \\ \hline
    0.5  & 25 & 30 & -5(-16.67\%)  \\ \hline
    0.75 & 25 & 30 & -5(-16.67\%)  \\ \hline
    1    & 25 & 30 & -5(-16.67\%)  \\ \hline
    2    & 22 & 21 &  1(4.76\%)    \\ \hline
    5    & 12 & 12 &  0(0.00\%)    \\ \hline
    10   & 9  & 9  &  0(0.00\%)    \\ \hline
    20   & 6  & 6  &  0(0.00\%)    \\ \hline
  \end{tabular}%
 }
 \caption{Derived radius of break-point of material using our approach, compared to the 'ground truth' obtained from the simulation under domain size of 30. The table presents the differences and corresponding error rates (\%) for the radius under various threshold conditions.}
 \label{tab:accuacy-1}
\end{table}

\subsection{Overhead of Material Deformations Analysis}
The overhead is measured by execution time of our approach, and the results are presented into Table \ref{tab:LULESH_overhead_1}.
No significant differences among the overheads using different MPI$\times$OpenMP configurations.
As the problem size increases to 60 and 90, the execution times for the original configurations increase significantly,  while the overhead remains lower than 3\% in terms of ratio.
The result indicates that while feature extraction does introduce some overhead, the impact is minimal across various configurations, making it a viable approach for enhancing simulation efficiency.

\begin{table*}[!ht]
  \centering
  \resizebox{\textwidth}{!}{
    \begin{tabular}{|p{0.15\linewidth}|p{0.08\linewidth}|p{0.08\linewidth}|p{0.1\linewidth}|p{0.08\linewidth}|p{0.08\linewidth}|p{0.1\linewidth}|p{0.08\linewidth}|p{0.08\linewidth}|p{0.1\linewidth}|}
    \hline
    \multirow{3}{0.15\linewidth}{\textbf{MPI$\times$OpenMP}} & \multicolumn{9}{c|}{\textbf{Problem Sizes}} \\ \cline{2-10}
    & \multicolumn{3}{p{0.32\linewidth}|}{30 $\times$ 30 $\times$ 30} & \multicolumn{3}{p{0.32\linewidth}|}{60 $\times$ 60 $\times$ 60} & \multicolumn{3}{p{0.32\linewidth}|}{90 $\times$ 90 $\times$ 90} \\ \cline{2-10}
    & origin & non-stop & overhead(\%) & origin & non-stop & overhead(\%) & origin & non-stop & overhead(\%) \\ \hline
    1$\times$1  & 7.2563 & 7.2803 & 0.024(0.33\%) & 69.667 & 70.667 & 1.0003(1.43\%)& 404.14 & 406.66 & 2.516(0.63\%) \\ \hline 
    8$\times$1  & 3.5447 & 3.637  & 0.0923(2.6\%) & 61.180 & 70.667 & 0.1958(0.32\%)& 304.67 & 305.31 & 0.6418(0.21\%) \\ \hline 
    27$\times$1 & 1.6725 & 1.6879 & 0.0154(0.92\%)& 20.599 & 20.992 & 0.393(1.91\%) & 116.78 & 117.03 & 0.245(0.21\%)\\ \hline 
    \end{tabular}
  }
  \caption{Execution time of LULESH, LULESH with feature extraction, and the corresponding overhead under various conditions of domain sizes, number of MPI processors, and number of OpenMP threads.}
  \label{tab:LULESH_overhead_1}
\end{table*}

We conducted experiments to evaluate how early termination can optimize our feature extraction process.
Table \ref{tab:LULESH_overhead_2} presents performance metrics, including the predicted radius, actual number of iterations, and execution time for identifying the break-point of material under various velocity thresholds expressed as percentages of the initial velocity at the point of contact.
Across all thresholds, the workload for these iterations appears to be evenly distributed.
For example, at thresholds of 0.1\% and 0.2\%, the number of iterations required to identify the region of interest consistently accounts for 40\% of the total iterations, with an equivalent execution time of approximately 41.64\% of the total execution time for the 30 problem size.
For the larger sizes of 60 and 90, the execution times for these 40\% of iterations are 41.53\% and 40.47\%, respectively.
These results underscore the effectiveness of our early termination approach in addressing the performance challenges inherent in hydrodynamics simulations.
By allowing the simulation to terminate once a predefined accuracy threshold is achieved, our method accelerates the overall simulation process while ensuring reliable identification of material break-points.
The findings also indicate that the choice of threshold significantly influences both the efficiency and effectiveness of the simulation, with optimal thresholds facilitating quicker identification of critical features and minimizing computational overhead.

\begin{table*}[!ht]
  \centering
  \resizebox{\textwidth}{!}{
    \begin{tabular}{|p{0.12\linewidth}|p{0.12\linewidth}|p{0.12\linewidth}|p{0.12\linewidth}|p{0.12\linewidth}|p{0.12\linewidth}|p{0.12\linewidth}|p{0.12\linewidth}|p{0.12\linewidth}|p{0.12\linewidth}|}
    \hline
    \multirow{3}{0.1\linewidth}{\textbf{Threshold (\% of the initial)}} & \multicolumn{9}{c|}{\textbf{Problem Sizes (number of iterations for full simulation)}} \\ \cline{2-10}
    & \multicolumn{3}{p{0.2\linewidth}|}{30 $\times$ 30 $\times$ 30(932)} & \multicolumn{3}{p{0.2\linewidth}|}{60 $\times$ 60 $\times$ 60(2031)} & \multicolumn{3}{p{0.2\linewidth}|}{90 $\times$ 90 $\times$ 90(3145)} \\ \cline{2-10}
    & Region radius & \# Iterations when region-of-interest is identified(\% of 932) & Execution time in seconds(\% of total execution time) & Region radius & \# Iterations when region-of-interest is identified(\% of 2031) & Execution time in seconds(\% of total execution time) & Region radius & \# Iterations when region-of-interest is identified(\% of 3145) & Execution time in seconds(\% of total execution time) \\ \hline
    0.1  & 30 & 373(40.0\%) & 3.0218(41.64\%) & 60 & 812(40.0\%) & 28.9368(41.53\%) & 90 & 1258(40.0\%) & 163.587(40.47\%) \\ \hline
    0.2  & 30 & 373(40.0\%) & 3.0218(41.64\%) & 60 & 812(40.0\%) & 28.9368(41.53\%) & 90 & 1258(40.0\%) & 163.587(40.47\%) \\ \hline
    0.5  & 30 & 373(40.0\%) & 3.0218(41.64\%) & 50 & 812(40.0\%) & 28.9368(41.53\%) & 51 & 1258(40.0\%) & 163.587(40.47\%) \\ \hline
    0.75 & 30 & 373(40.0\%) & 3.0218(41.64\%) & 42 & 812(40.0\%) & 28.9368(41.53\%) & 42 & 1258(40.0\%) & 163.587(40.47\%) \\ \hline
    1    & 30 & 373(40.0\%) & 3.0218(41.64\%) & 34 & 812(40.0\%) & 28.9368(41.53\%) & 34 & 1258(40.0\%) & 163.587(40.47\%) \\ \hline
    2    & 21 & 373(40.0\%) & 3.0218(41.64\%) & 22 & 406(20.0\%) & 14.5633(20.90\%) & 22 & 629 (20.0\%) & 81.5740(20.18\%) \\ \hline
    5    & 12 & 373(40.0\%) & 3.0218(41.64\%) & 13 & 406(20.0\%) & 14.5633(20.90\%) & 13 & 629 (20.0\%) & 81.5740(20.18\%) \\ \hline
    10   & 9  & 373(40.0\%) & 3.0218(41.64\%) & 9  & 406(20.0\%) & 14.5633(20.90\%) & 9  & 629 (20.0\%) & 81.5740(20.18\%) \\ \hline
    20   & 6  & 373(40.0\%) & 3.0218(41.64\%) & 6  & 406(20.0\%) & 14.5633(20.90\%) & 6  & 629 (20.0\%) & 81.5740(20.18\%) \\ \hline
    \end{tabular}
  }
  \caption{Performance of LULESH for identifying the break-point of the material under various thresholds, specified as percentages of the initial velocity at the point of contact. The table includes the absolute iteration counts required to locate regions of interest, as well as these counts as percentages of the total iterations in a full simulation. Additionally, the execution time for simulations terminated at each threshold is provided, along with their proportions relative to the total simulation execution time.}
  \label{tab:LULESH_overhead_2}
  \vspace{-0.3cm}
\end{table*}

\section{Case 2: WD Mergers Detonation Determination}
\label{sec:castro}
Type Ia supernovae (SNe Ia) are renowned for their extreme luminosity, making them essential tools for cosmological distance measurement~\cite{perlmutter2003measuring}.
These events result from thermonuclear explosions of WDs~\cite{kilic2010elm}, and understanding the triggers of WD detonation is a significant area of study~\cite{zingale2018meeting}.
The delay-time distribution (DTD), which represents the supernova rate over time following an initial star formation burst~\cite{totani2008delay}, provides key insights into explosion mechanisms and connects theoretical predictions with observations~\cite{mennekens2010delay}.
Since DTD is closely linked to stellar and binary evolution timescales, different progenitor scenarios result in distinct DTDs~\cite{maoz2012delay}, adding complexity to its analysis.

\begin{figure}[!htb]
     \centering
     \begin{subfigure}{0.15\textwidth}
         \centering
         \includegraphics[width=\textwidth]{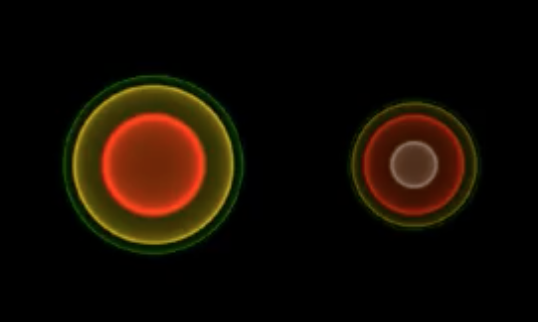}
         \caption{Initial}
         \label{fig:e1}
     \end{subfigure}
     \hfill
     \begin{subfigure}{0.15\textwidth}
         \centering
         \includegraphics[width=\textwidth]{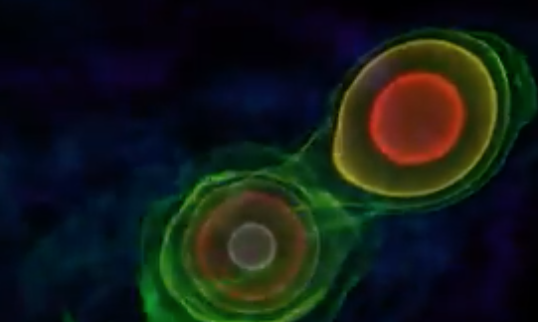}
         \caption{Merging}
         \label{fig:e2}
     \end{subfigure}
     \hfill
     \begin{subfigure}{0.15\textwidth}
         \centering
         \includegraphics[width=\textwidth]{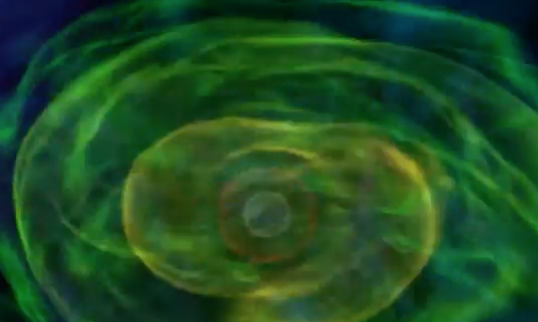}
         \caption{Boiling}
         \label{fig:e3}
     \end{subfigure}
        \caption{Illustration of WD merger, adapted from ~\cite{katz2016white}. It starts from a system including a primary and a secondary star, continue to merge, and become boiling thus to be detonated by collapse or violent merging.}
        \label{fig:wdmerger}
\end{figure}

The Castro wdmerger simulation~\cite{almgren2020castro} models SNe Ia through WD mergers, where a primary WD accretes material from a secondary WD until it approaches the Chandrasekhar mass limit. An illustration adapted from~\cite{katz2016white} (Figure \ref{fig:wdmerger}) shows the evolution from a binary system to a merger, leading to detonation through collapse or violent merging.
By integrating gravitational and rotational forces with hydrodynamics, the simulation effectively captures mass advection and system dynamics over extended timescales~\cite{katz2016white}.

We apply our approach to the Castro wdmerger simulation to demonstrate its capacity for efficiently tracking the evolution of various astrophysical events, thereby providing insights into the DTD within the WD merger progenitor scenario.
In our experiment, we extract four key features, including temperature, angular momentum, mass, and energy distributions.
The temporal and spatial trends on the area crossing origin of the domain are obtained by curve-fitting, and the features are analyzed by variable tracking.
The inflection point are the one to be tracked.
It is recognized as the signal of stage changes of these variables, thus work as the indicator of thermonuclear detonation.
For example, the inflection point of mass, which is the connection of plateau and decreasing part, is the signal of material ejection showing a potential detonation of the remnant.
As a result, a delay-time is derived corresponding to each identified detonation event from the extracted features, providing a comprehensive timeline of supernova activity and deeper insights into the evolution of these explosive phenomena.

We execute the application for evaluation under conditions of 3D domain resolutions of 16, 32, and 48, number of MPI processors of 8, 16, and 32, and number of OpenMP threads of 1, 2, and 4.
The accuracy is evaluated by comparing the delay-time derived from our extracted features with that obtained from the complete simulation dataset, which serves as the 'ground truth'.
The overhead is assessed by measuring the difference in execution time between simulations with and without our approach, expressed as a percentage of the total simulation runtime.
To demonstrate the potential of our approach for computational efficiency, we performed curve-fitting on datasets collected at different simulation stages, specifically at 10\%, 25\%, and 50\% of total iterations.
By enabling early termination of the simulation once the auto-regressive model reached a predefined accuracy threshold, our approach showed substantial promise in reducing overall computational workload while maintaining reliable feature extraction.

\subsection{Accuracy of WD Mergers Detonation Determination}
\begin{figure}[!htb]
     \centering
     \begin{subfigure}{0.23\textwidth}
         \centering
         \includegraphics[width=\textwidth]{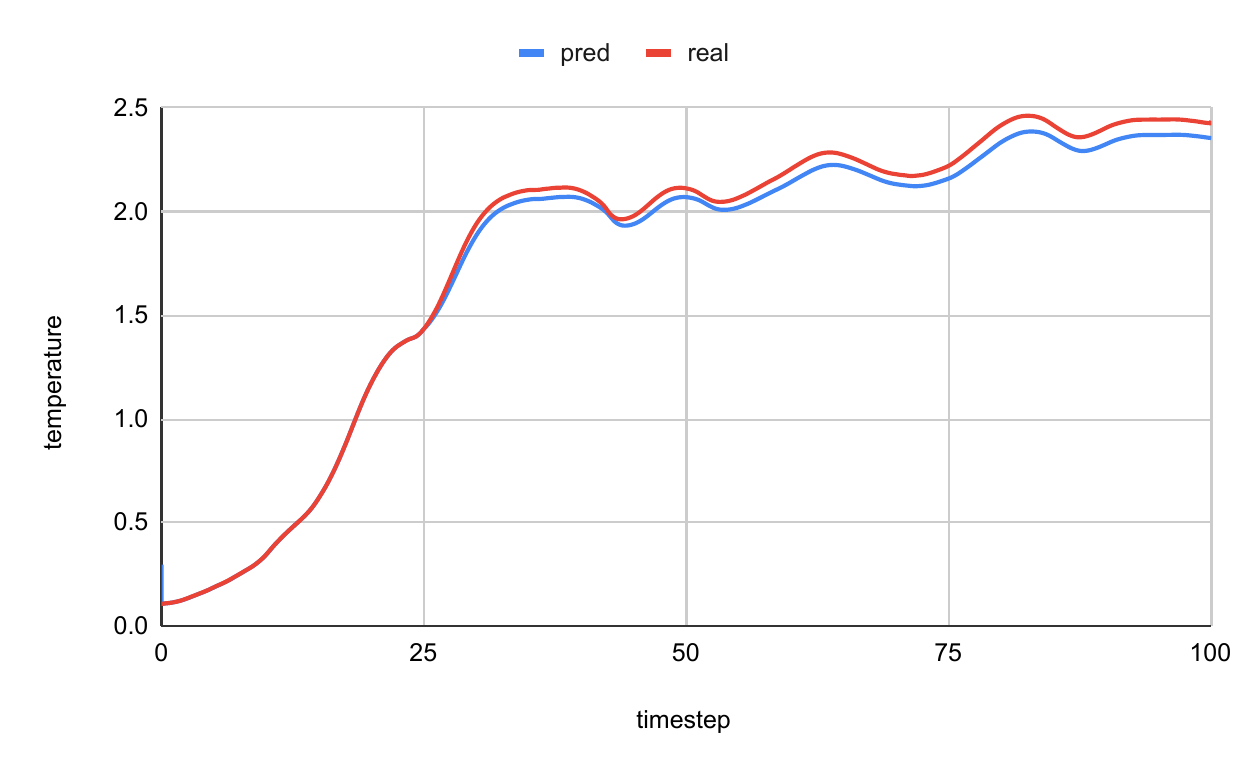}
         \caption{Temperature}
         \label{fig:temp}
     \end{subfigure}
     \hfill
     \begin{subfigure}{0.23\textwidth}
         \centering
         \includegraphics[width=\textwidth]{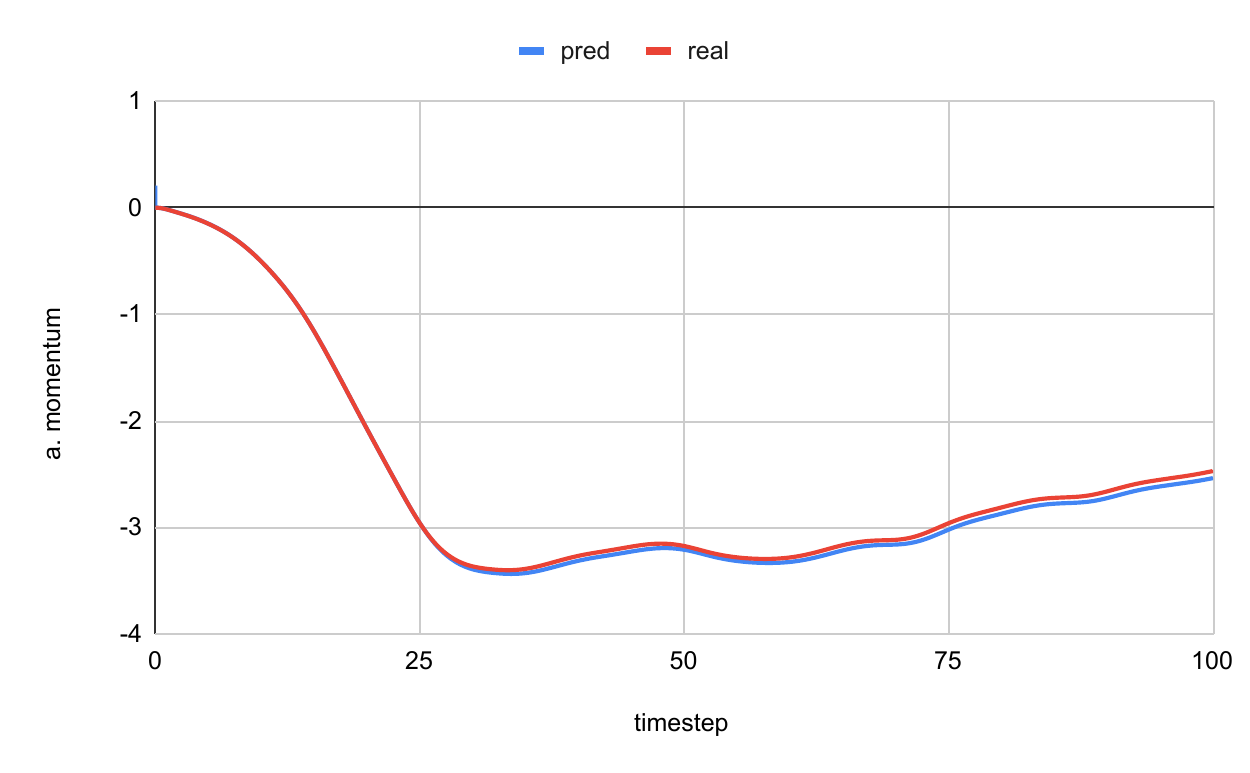}
         \caption{A. Momentum}
         \label{fig:mom}
     \end{subfigure}
     \hfill
     \begin{subfigure}{0.23\textwidth}
         \centering
         \includegraphics[width=\textwidth]{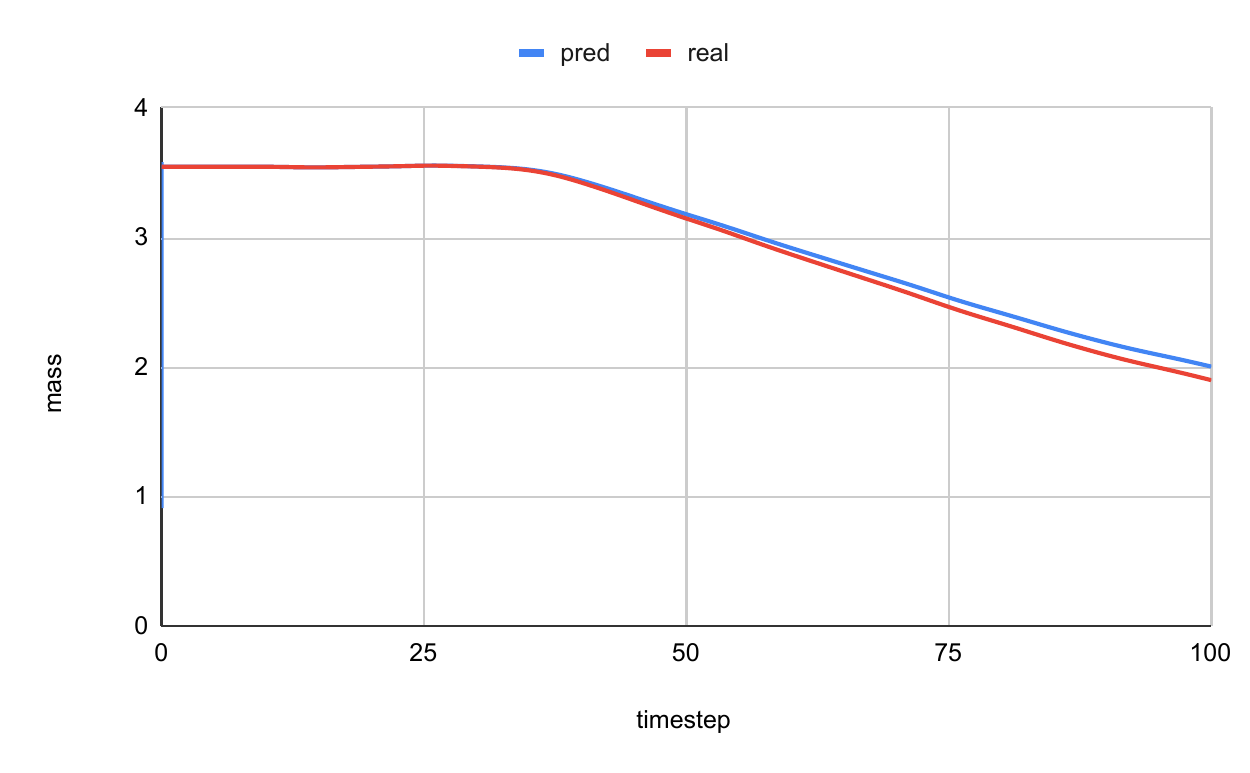}
         \caption{Mass}
         \label{fig:mass}
     \end{subfigure}
     \hfill
     \begin{subfigure}{0.23\textwidth}
         \centering
         \includegraphics[width=\textwidth]{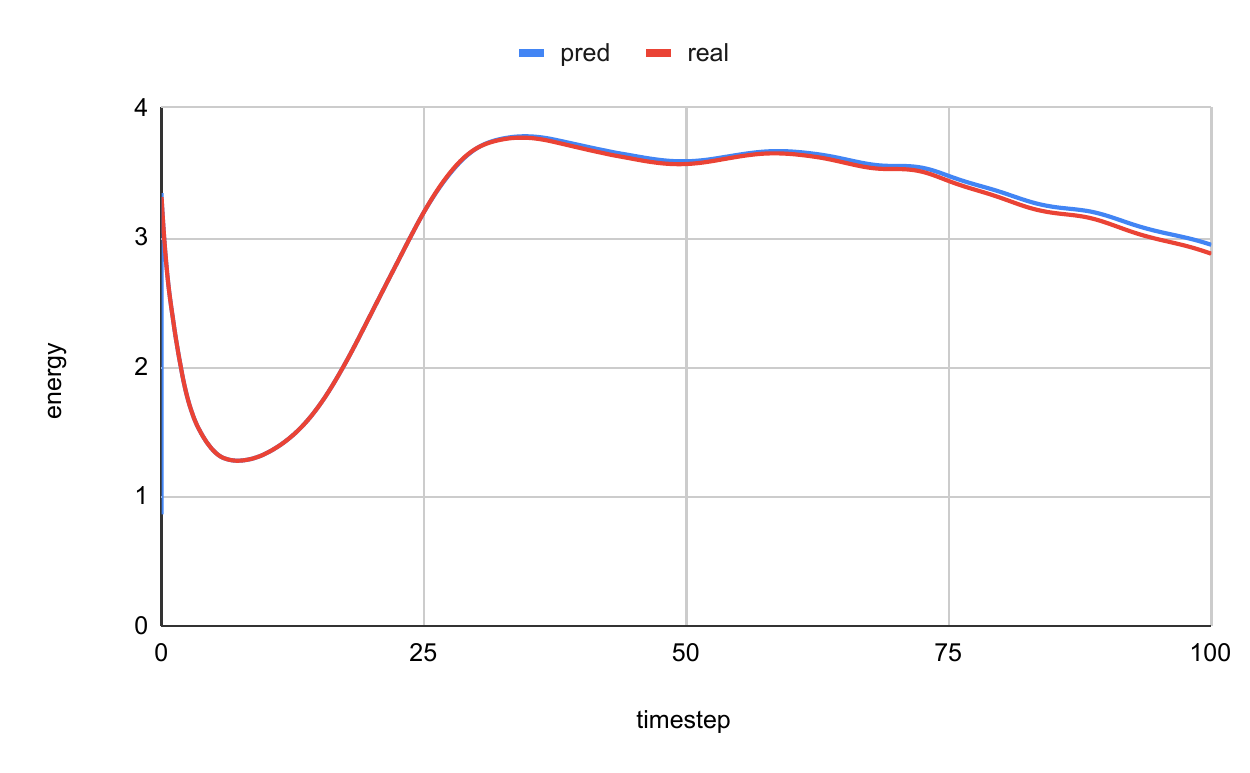}
         \caption{Energy}
         \label{fig:energy}
     \end{subfigure}
        \caption{Curve-fitting results on various diagnostic variables using 25\% of total iterations for training. The crosponding data directly generated from simulation are also visualized for comparison.}
        \label{fig:pred_real}
\end{figure}

The accuracy of curve fittings for various diagnostic variables, including temperature, angular momentum, mass, and energy, is presented in Table \ref{tab:pred}.
The training data were obtained from simulations conducted on a 3D domain with a resolution of 32, incorporating 10\%, 25\%, and 50\% of the total iterations.
According to Table \ref{tab:pred}, we observed that the accuracy is increasing when the number of training iterations is increased.
Among these diagnostic variables, we also observed that the overall accuracy of mass is not strongly impacted by the volume of training data, while the others like temperature, angular momentum, and energy are significantly effected.
An illustration of the curve-fitting on all diagnostic variables under 25\% of total iterations is shown in Figure \ref{fig:pred_real}.
To maintain high accuracy and benefit from workload reduction via early termination of simulation, a ratio of 25\% is chosen.

\begin{table}[!htb]
\resizebox{\linewidth}{!}{%
  \begin{tabular}{|p{0.3\linewidth}|p{0.2\linewidth}|p{0.2\linewidth}|p{0.2\linewidth}|}
  \hline
  \multirow{2}{0.3\linewidth}{\textbf{Diagnostic Var.}} & \multicolumn{3}{|c|}{\textbf{Ratio of Total Number of Iterations (\%)}} \\ \cline{2-4}
  & \textbf{10} & \textbf{25} & \textbf{50} \\ \hline
  Temperature   & 18.6\% & 2.7\%  &  0.69\% \\ \hline
  A. Momentum   & 9.92\% & 1.81\% &  0.69\% \\ \hline
  Mass          & 3.05\% & 3.08\% &  2.02\% \\ \hline
  Energy        & 8.3\%  & 1.16\% &  0.56\% \\ \hline
  \end{tabular}%
}
\caption{Error rates of curve-fitting (\%) for various diagnostic variables, using training data from 10\%, 25\%, and 50\% of the total iterations under a domain resolution of 32.}
\label{tab:pred}
\end{table}


\begin{figure}[h]
\centering
  \includegraphics[width=0.8\linewidth]{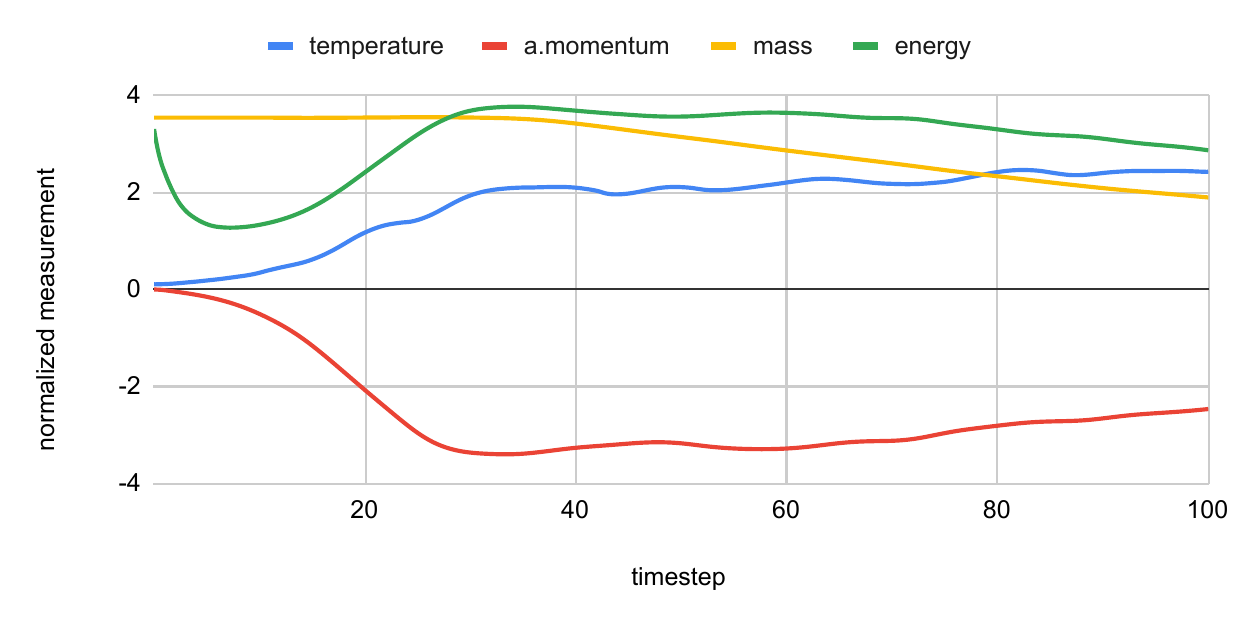}
  \caption{Distribution of diagnostic variables of temperature, angular momentum, mass, and energy over timesteps. The inflection points of these lines are indicators of denotations.}
  \label{fig:tracking}
\end{figure}

We applied our approach to Castro wdmerger to obtain the distribution of key variables over timesteps, which is the measurement of physical reaction time elapsed in terms of simulation of merging WDs.
An example of the curves of diagnostic variables including temperature, angular momentum, mass, and energy over timesteps plotted with data collected from Castro wdmerger is shown in Figure \ref{fig:tracking} and the derived delay times are shown in Table \ref{tab:accuacy-1}.

\begin{table}[!htb]
 \resizebox{\linewidth}{!}{%
  \begin{tabular}{|p{0.3\linewidth}|p{0.3\linewidth}|p{0.3\linewidth}|p{0.3\linewidth}|}
  \hline
    \textbf{Diagnostic Var.} & \textbf{From Sim.} & \textbf{Feat. Extraction} & \textbf{Difference(\%)}\\ \hline
    Temperature   & 31.236 & 30.8394 & -0.3972(-1.27\%) \\ \hline
    A. Momentum   & 32.436 & 30.3085 & -2.1276(-6.56\%) \\ \hline
    Mass          & 31.363 & 31.2306 & -0.1265(-0.4\%) \\ \hline
    Energy        & 30.707 & 32.1669 &  1.4599(4.75\%) \\
  \hline
  \end{tabular}%
 }
 \caption{Derived delay-time of thermonuclear detonation using our approach, compared to the 'ground truth' obtained from the simulation. The table presents the differences and corresponding error rates (\%) for the delay time obtained via various diagnostic variables.}
 \label{tab:accuacy-1}
\end{table}

We observed a collection of inflection points being closely aligned to the delay-time of 30. These inflection points, such as a slowdown increment of temperature and energy, a slowdown decreasing of angular momentum, and a decreasing of mass are all signals of the potential of thermonuclear detonation. 
For diagnostic variables of temperature, angular momentum, mass, and energy, the rate of increase in its value suddenly decreases.
To adapt this behavior to our method, we note that the gradient of the time-scale ratio quickly drops.
By comparing the gradient of this timestamp with those of the preceding and following timesteps, a delay time can thus be derived. 
According to Table \ref{tab:accuacy-1}, we also observed that due to the accuracy of simulation prediction under 25\% of iterations, the final derived delay-time of thermonuclear detonation are of small difference, which can be as small as 0.4\%.

\subsection{Overhead of WD Mergers Detonation Determination}

\begin{table*}[!htb]
\centering
\resizebox{\linewidth}{!}{%
\begin{tabular}{|c|c|c|c|c|c|c|c|c|c|c|c|c|c|c|c|}
  \hline
  \multirow{3}{*}{\textbf{MPI × OpenMP}} & \multicolumn{15}{c|}{\textbf{Domain Resolutions}} \\ \cline{2-16}
  & \multicolumn{5}{c|}{$16 \times 16 \times 16$} & \multicolumn{5}{c|}{$32 \times 32 \times 32$} & \multicolumn{5}{c|}{$48 \times 48 \times 48$} \\ \cline{2-16}
  & Orig & No-stop & Ovh & Stop & Acc & Orig & No-stop & Ovh & Stop & Acc & Orig & No-stop & Ovh & Stop & Acc \\ \hline
  $8 \times 1$  & 83.19 & 84.52 & 1.61\% & 42.15 & 49.3\%  & 151.84 & 153.53 & 1.11\% & 61.16 & 59.7\%  & 1475.69 & 1478.06 & 0.16\% & 485.51 & 67.1\%  \\ \hline
  $8 \times 2$  & 70.70 & 72.69 & 2.82\% & 36.27 & 48.7\%  & 132.65 & 132.71 & 0.05\% & 53.54 & 59.6\%  & 793.62  & 795.05  & 0.18\% & 267.42 & 66.3\%  \\ \hline
  $8 \times 4$  & 56.27 & 57.59 & 2.34\% & 29.02 & 48.4\%  & 108.50 & 110.69 & 2.02\% & 44.19 & 59.2\%  & 460.49  & 462.82  & 0.51\% & 155.84 & 66.2\%  \\ \hline
  $16 \times 1$ & 83.85 & 85.49 & 1.95\% & 41.22 & 50.8\%  & 153.52 & 155.14 & 1.06\% & 61.68 & 59.8\%  & 467.07  & 469.11  & 0.44\% & 156.36 & 66.5\%  \\ \hline
  $16 \times 2$ & 72.20 & 73.04 & 1.16\% & 36.36 & 49.6\%  & 132.47 & 135.65 & 2.41\% & 53.34 & 59.7\%  & 353.62  & 355.60  & 0.56\% & 118.49 & 66.4\%  \\ \hline
  $32 \times 1$ & 89.63 & 91.66 & 2.26\% & 45.88 & 48.8\%  & 162.82 & 164.13 & 0.80\% & 65.50 & 59.7\%  & 275.88  & 278.66  & 1.01\% & 93.34  & 66.1\%  \\ \hline
\end{tabular}%
}
\caption{Execution time (Orig), with feature extraction (No-stop), with early termination (Stop), and the corresponding overhead (Ovh) and acceleration (Acc) under various MPI × OpenMP configurations and domain sizes.}
\label{tab:overhead-2}
\vspace{-0.3cm}
\end{table*}

To assess the overhead of our method, we recorded the execution time of the simulation both with and without feature extraction. To gain deeper insights into performance related to execution time, we further examined the potential for accelerating the simulation through early termination for each domain size.
For the potential of acceleration, according to the rule of early termination mentioned before, we run the simulation with feature extraction and terminate the simulation when the auto-regression model is well trained.
We compute the differences as well as the ratio between the execution time of original simulation and the simulation with feature extraction and early-termination.
The results are recorded in Table \ref{tab:overhead-2}.
According to Table \ref{tab:overhead-2}, the overhead associated with simulation prediction varies from 0.05\% to 2.82\%.
This indicates that our feature extraction method has minimal impact on overall simulation performance.
Furthermore, we observed a significant potential for acceleration, achieving a reduction ratio of up to 67.1\% in cases with a domain resolution of 48.
This suggests that early termination of the simulation can effectively streamline processing while maintaining the accuracy of the results.

Based on the discussion above, we found that the error rate for delay-time estimates ranges from 0.4\% to 6.5\%, while the overhead associated with executing feature extraction varies from 0.05\% to 2.82\%.
We conclude that our real-time, auto-regression-based in-situ method effectively balances accuracy and overhead in determining detonation events in WD mergers.
Moreover, our approach extends beyond simply tracking detonations in WD mergers; it demonstrates significant potential for optimizing threshold detection tasks by allowing for early termination of simulations once the threshold is reached. Our experiments indicate that this early-termination strategy can accelerate total execution time by up to 67.1\%.
In addition, our method provides critical data points for the delay time of detonations, contributing to the reconstruction of DTDs from WD merger-based progenitor systems.
This dynamic solution for analyzing DTDs offers advantages over methods that rely on fixed timescales, such as those proposed by Tutukov et al.~\cite{tutukov1994merging} and de Donder et al.~\cite{de2004influence}.

\section{Related Work}
\label{sec:related_work}
Traditional feature extraction methods primarily focus on accuracy, often involving large amount of data to enhance precision.
For instance, Corsico et al.\cite{corsico2001potential} incorporate mass alongside luminosity variations to analyze the physical properties of white dwarf (WD) mergers, while Gubenko et al.\cite{gubenko2008determination} use density as a supplement to temperature profiling, improving analysis for large-radius WDs~\cite{vikhlinin2006chandra}.
Additionally, the choice of data analysis models plays a crucial role in achieving high accuracy.
Beyond leveraging complex models to maximize data insights~\cite{zhang2022prediction}, scientists often apply calibration and verification techniques to ensure precise feature extraction~\cite{cheng1991interfacing}.
However, these accuracy-driven approaches typically involve trade-offs, as they may compromise computational performance.

In-situ methods are designed to address the significant I/O challenges often faced by approaches that rely on data transport~\cite{dutta2021situ}, allowing for efficient 'real-time' data analysis in various scenarios.
For example, Li et al.\cite{li2017real} introduce a machine learning (ML) method that provides real-time predictions of turbulence flow evolution, while Amatore et al.\cite{amatore2005situ} propose a threshold-based method for monitoring flow profiles within a microfluidic channel in real time.
To minimize disruptions to active simulations, the impact of in-situ methods on simulation performance must be carefully managed.
However, they can still introduce delays, especially during intensive data processing tasks such as large-scale regression~\cite{ayachit2016performance}, image rendering~\cite{larsen2016performance}, or interactions involving human input~\cite{yi2014situ}.

To minimize the impact on simulation performance, feature extraction methods utilize reduced datasets during simulations, aiming to capture a representative scope of the entire dataset~\cite{bryson1999visually}.
For example, Silva et al.\cite{silva2017raw} present an approach that extracts subsets of data generated from computational fluid dynamics (CFD) simulations, focusing on those subsets that exhibit strong connections to hydrodynamic features.
In cases where these relationships are complex and difficult to extract directly, one-dimensional data analysis is employed.
For example, Cao et al.\cite{cao2022digital} achieve real-time monitoring of tidal turbines by concentrating exclusively on spatial data related to horizontal axis turbine performance.
Similarly, Bryson et al.~\cite{bryson1999visually} visualize streamlines around aircraft in real time by rendering temporal data along specific time steps.
However, despite enabling real-time analysis, these methods still face challenges in balancing accuracy with computational overhead.

To make it easier for users to implement data analysis, library frameworks are presented to enhance the programming productivity.
For example, Dutta et al.~\cite{dutta2022situ} integrate customized functions for accessing to different types of simulation data to AMReX.
The data is saved for later analysis as well as visualization.
ParaView~\cite{ayachit2015paraview} and DfAnalyzer~\cite{silva2018dfanalyzer} are designed for enhanced real-time analysis, and can be applied to wide range of applications.
Besides the universal ones, library frameworks or software for a specific domain are designed.
For example, Stone et al.~\cite{stone2009wavcis} present various models for data analysis for ocean models exclusively.
They are of complex design, and they are not well suited for quick start of in-situ analysis.

\section{Conclusion}
\label{sec:conclusion}
In this paper, we introduced a real-time, auto-regression-based approach for in-situ data analysis in hydrodynamics simulations, along with a flexible library framework enabling users to effectively integrate our method into their workflows.
Through evaluations on material deformation analysis and WD detonation determination, our feature extraction method provided valuable insights, capturing key behaviors across both simulations.
Additionally, by allowing for early termination of the simulation once the auto-regressive model reached sufficient accuracy, our approach demonstrated significant potential for reducing computational workload.
In conclusion, our method achieves accurate feature extraction while adding minimal overhead to the simulation process.


\bibliographystyle{IEEEtran}
\bibliography{references}

\end{document}